\documentclass[11pt]{article}

\usepackage{amsmath}
\usepackage{amsfonts}
\usepackage{amssymb}
\usepackage{latexsym}
\usepackage{color}
\usepackage[dvipdfmx]{graphicx}
\usepackage{cite}
\usepackage{subfigure}
\input{colordvi.tex}

\begin{document}

\begin{titlepage}
\begin{flushright}
RESCEU-21/15

\end{flushright}
\begin{center}


\vskip .5in

{\Large \bf
Spontaneous scalarization: asymmetron as dark matter
}
\vskip .45in

{
Pisin Chen\footnote{This work was initiated and largely advanced while the author was staying at 
RESCEU as a visiting professor. }$^{a,b,c}$,
Teruaki Suyama$^{a}$
and Jun'ichi Yokoyama$^{a,d,e}$
}


{\em
$^a$
   Research Center for the Early Universe (RESCEU), Graduate School
  of Science,\\ The University of Tokyo, Tokyo 113-0033, Japan
}\\
{\em
$^b$
   Department of Physics and Graduate Institute of Astrophysics
   \& Leung Center for Cosmology and Particle Astrophysics, 
   National Taiwan University, Taipei, Taiwan 10617
}\\
{\em
$^c$
  Kavli Institute for Particle Astrophysics and Cosmology,\\
  SLAC National Accelerator Laboratory, Stanford University, Stanford, CA 94305, U.S.A.
}\\
{\em
$^d$
  Department of Physics, Graduate School of Science,\\ The University of Tokyo, Tokyo 113-0033, Japan
}\\
{\em
$^e$
  Kavli Institute for the Physics and Mathematics of the Universe, WPI, TODIAS,\\
  The University of Tokyo, Kashiwa, Chiba 277-8568, Japan
}
\\

\end{center}

\vskip .4in

\begin{abstract}
We propose a new scalar-tensor model which induces significant deviation from
general relativity inside dense objects like neutron stars, while passing
solar-system and terrestrial experiments, extending a model proposed
by Damour and Esposito-Farese. Unlike their model, we employ a massive
scalar field dubbed asymmetron so that it not only realizes proper cosmic
evolution but also can account for the cold dark matter.
In our model, asymmetron undergoes spontaneous scalarization inside 
dense objects, which results in reduction of the gravitational constant
by a factor of order unity.
This suggests that observational tests of constancy of the 
gravitational constant in high density phase are the effective
ways to look into the asymmetron model.
\end{abstract}
\end{titlepage}

\renewcommand{\thepage}{\arabic{page}}
\setcounter{page}{1}
\renewcommand{\thefootnote}{\#\arabic{footnote}}

\section{Introduction}
General relativity(GR), which describes gravity in terms of a massless
spin 2 field, has been tested and passed all the 
precision experimental tests such as the solar system
and the terrestrial experiments \cite{Will:2014kxa}. 
No observations which clearly contradict with predictions of
GR have been found.
This does not guarantee that GR remains valid under extreme physical conditions beyond the present experimental limits.
Indeed, it is known that GR cannot tell what happens at the center of
black holes and at the very beginning of the Universe if energy conditions
are to be satisfied \cite{Hawking-Ellis}.
In addition to this, the explanation of the accelerating expansion of 
the late time universe may require the modification of GR on very large scales.
Motivated by these considerations, GR may be viewed as an effective theory 
which is valid only in some domain of space spanned by physical parameters 
such as length, energy and density,
although the boundary of such domain is not yet well defined.
Many possibilities have been proposed in the literature in various 
contexts (see references in \cite{DeFelice:2010aj,deRham:2014zqa,Joyce:2014kja,Berti:2015itd}).

In the near future, direct detection of gravitational waves will become possible
by using the laser interferometers such as advanced Laser 
Interferometric Gravitationalwave Observatory (aLIGO)\cite{Harry:2010zz}, 
advanced Virgo (aVirgo) \cite{TheVirgo:2014hva} and KAGRA \cite{Aso:2013eba}.
Target gravitational waves originate from the vicinity of the compact objects
such as neutron stars where the matter density is much larger than any other
places in the Universe.
Observation and analysis of such gravitational waves should enable us
to probe the laws of gravity in such domain 
that has heretofore unexplored.

Scalar-tensor(ST) theories are well-studied and natural alternatives to 
GR \cite{Bergmann:1968ve,Nordtvedt:1970uv,Wagoner:1970vr,Damour:1992we}.
Observations of gravitational waves enable us to probe ST theories 
in high density and strong gravity regime.
Interesting targets relevant to gravitational wave observations
are a class of ST theories which mimics GR in low density (or weak gravity) 
regime but significantly deviates from GR in 
high density(or strong gravity) regime \cite{Harada:1998ge,Pani:2010vc}.
One natural way to construct such a model is to introduce interaction
between the standard model particles and the scalar field by the conformal factor
so that the effective potential for the scalar field depends on the matter density.
If the system is static, the scalar field takes a value that minimizes 
the energy of the system.
This expectation value depends on the matter density and controls 
the interaction strength between the standard model particles \cite{Khoury:2003aq}.
Then, it is possible that the expectation value vanishes if matter density is 
low and the spontaneous scalarization occurs if matter density exceeds a critical value.
In such a case, modification of GR occurs only in the 
high density region exceeding the critical density.
This mechanism is completely opposite to the symmetron model proposed in \cite{Hinterbichler:2010es}
in which scalarization occurs only when the matter density 
becomes smaller than the critical density.
In that model, matter density inside the solar system is supposed to
be larger than the critical density and GR is recovered but
deviation from GR appears on cosmological environment due to the low 
background density.
For this reason, we call the scalar field that acquires non-vanishing
expectation value only in high density environments {\it asymmetron}.

\begin{figure}[tbp]
  \begin{center}
   \includegraphics[width=130mm]{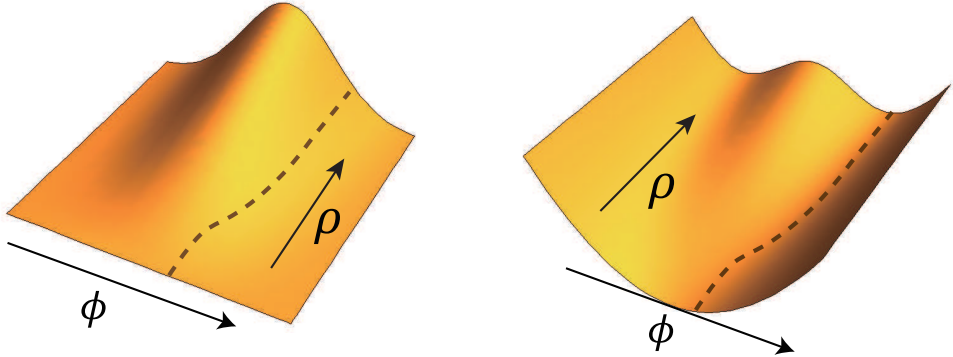}
  \end{center}
  \caption{Density dependent effective potential of the scalar field for the 
  Damour-Esposito-Farese model (left figure) and the asymmetron model 
  we consider in this paper (right figure). 
  Dotted curve in each figure represents the value of $\phi$ that
  $\phi$ would take inside the compact object with density $\rho$. }
  \label{fig:EP}
\end{figure}

In this context, there is an interesting scalar-tensor theory 
proposed by Damour and Esposito-Farese (DEF) \cite{Damour:1993hw,Damour:1996ke}
in which significant deviation from GR occurs only 
in the vicinity and the inside of 
neutron stars and safely passes the
solar system experiments.
In the DEF model, the scalar field in high matter density region becomes tachyonic due to
a particular form of the conformal coupling with the standard model particles 
(see left figure of Fig.~\ref{fig:EP}).
As a result, the scalar field takes a large non-vanishing value inside the neutron star and
approaches a non-vanishing but much smaller value at distance far away from the star.
The value at infinity is fixed to match
the cosmological value just as in the case of the Fierz-Jordan-Brans-Dicke theory
\cite{Jordan:1949zz,Fierz:1956zz,Jordan:1959eg,Brans:1961sx}
and this value must be small enough to satisfy the solar system 
and terrestrial observational constraints.
Since the magnitude of the scalar field controls the amount of deviation from GR, 
significant deviation from GR occurs only in the inside or the vicinity of the neutron stars.
Because of this, the structure of the neutron stars differs from that under GR and
this suggests that studying neutron stars and deriving observational consequences 
is the most effective way to test DEF model, as has been pursued in the literature 
\cite{Barausse:2012da,Doneva:2013qva,Shibata:2013pra,Palenzuela:2013hsa,Sampson:2014qqa,Sotani:2014tua,Pani:2014jra,Taniguchi:2014fqa,Silva:2014ora,Ponce:2014hha}.

However, it is known that the DEF model faces difficulty of embedding it
in the cosmic history \cite{Damour:1992kf,Damour:1993id,Sampson:2014qqa}.
During inflationary and matter-dominated epochs, the coupling
between the scalar field and the matter field forces the scalar field
to take non-vanishing value
and the law of gravity in the present
Universe deviates from GR to an extent incompatible with the existing
constraints.
Our main motivation of this paper is to extend the DEF model to incorporate
it in the cosmological context.
We achieve this by dropping the two restrictions imposed in the DEF model.
The first is the mass of the scalar field and the second is the energy scale 
appearing in the conformal factor.
In the DEF model, the scalar field is assumed to be exactly massless
and the energy scale in the conformal factor is taken to be around the Planck mass.
We do not impose these conditions and assume that the scalar field
is massive and the energy scale in the conformal factor differs from the Planck scale.
Because of these assumptions, the effective potential of the
asymmetron has a global minimum for any value of 
the matter density $\rho$, whereas the effective potential of 
the original DEF model does not have such property.
(see Fig.~\ref{fig:EP}).

Let us first briefly explain how the extended DEF model can be consistently
embedded in the cosmology before describing quantitative analysis 
in the subsequent sections.
As is the case with the original DEF model,
in the extended DEF model the scalar field at the origin in the presence of matter
becomes unstable and should in principle be pushed away from the origin.
Thanks to the mass term, there exists a global minimum of the effective
potential, which helps the asymmetron to settle down at this point.
Assuming the universal conformal coupling, the scalarization should happen during inflation.
Due to the non-vanishing value of the asymmetron,
the gravitational constant would be different from the one we measure in the 
laboratory and in this sense the law of gravity would be different from GR 
as we know.
After inflation, the Universe is reheated and dominated by radiation.
Since the trace of the radiation energy-momentum tensor is zero,
asymmetron decouples from the matter and 
the global minimum shifts back to the origin of the effective potential.
As the Universe further expands, the Hubble parameter gradually decreases and
eventually becomes smaller than the mass of the asymmetron.
By then, the asymmetron undergoes damped oscillation and the Universe
gradually approaches GR.
That is, GR is a cosmological attractor in this model. 
As a result, GR is recovered to a good approximation in the present Universe.
We will further show that osillating component of the
asymmetron, which interacts only gravitationally with standard model
particles, is a good candidate for cold dark matter.
Therefore not only is our extended DEF model cosmologically viable but 
also it provides a dynamical mechanism for dark matter generation 
via asymmetron production during inflation.

Of course, cosmology is not the only arena where the extended DEF
model becomes relevant to observations.
When the matter density inside a compact astrophysical object 
exceeds the critical density, asymmetron would undergo the 
spontaneous scalarization and the laws of gravity may deviate
from GR by a considerable amount.
This phenomenon itself is similar to the original DEF model,
but our asymmetron model provides additional new features as follows.
Due to the mass term, asymmetron outside the compact object 
where no matter exists diminishes exponentially on the length scale of 
the inverse of the mass.
This is in clear contrast to the original DEF model where the 
asymptotic value is arbitrary and fixed by the boundary condition.
Furthermore, the critical density beyond which the spontaneous
scalarization occurs is not necessarily around the matter density of the
neutron star (it could be either higher or lower),
whereas in the original DEF model the spontaneous scalarization 
occurs inside the neutron star where the gravitational energy becomes 
comparable to its rest mass energy.
This opens up a new possibility that not only neutron stars
but also less compact astrophysical objects are the best targets to 
search for the deviation from GR.

In this paper, we analyze in detail the spontaneous scalarization
in the asymmetron model and how the gravitational field
changes outside the compact star before and after the spontaneous
scalarization. 
We also show that inflation, assuming the universal conformal
coupling to all the matter fields, induces the spontaneous
scalarization and the asymmetron undergoes coherent
oscillations in later time Universe. 
As mentioned above, such an oscillating field can be a candidate 
for cold dark matter.
We show that there is a parameter space where 
the production of asymmetron can saturate the dark matter content.
In the last section, we further comment on the possibility of the asymmetron
as dark energy.

\section{Spontaneous scalarization in high density region}
\subsection{Model}
We introduce a real massive scalar field $\phi$ which is universally coupled
to all the matter fields including the standard model particles 
through the metric ${\tilde g}_{\mu \nu}=A^2(\phi) g_{\mu \nu}$
(Thus, ${\tilde g}_{\mu \nu}$ is the Jordan metric).
This ensures that weak equivalence principle is satisfied.
We assume that $g_{\mu \nu}$ satisfies the Einstein equations.
Therefore, the basic action is given by
\begin{eqnarray}
S&=&S_g[g_{\mu \nu}]+S_\phi[g_{\mu \nu},\phi]+S_m[{\tilde g}_{\mu \nu},\phi] \nonumber \\
&=&\int d^4x \sqrt{-g} \left( \frac{R}{16\pi G_N}-\frac{1}{2} g^{\mu \nu} 
\partial_\mu \phi \partial_\nu \phi-\frac{\mu^2}{2}\phi^2 \right)+
\int d^4x \sqrt{-{\tilde g}} {\cal L}_m ({\tilde g},\psi_m),
\end{eqnarray}
where $G_N$ is the Newton's constant and ${\cal L}_m$ is the matter Lagrangian
of all the matter fields including the standard model fields.
The corresponding equations of motion are given by
\begin{eqnarray}
&&\Box \phi-\mu^2 \phi+A^3(\phi) A_{,\phi} {\tilde T}=0, \label{eq-phi} \\
&&G_{\mu \nu}=8\pi G_N \bigg[ -\left( \frac{1}{2} g^{\mu \nu} 
\partial_\mu \phi \partial_\nu \phi+\frac{\mu^2}{2}\phi^2 \right) g_{\mu \nu}+\partial_\mu \phi
\partial_\nu \phi+A^2(\phi) {\tilde T}_{\mu \nu} \bigg], \label{eq-g}
\end{eqnarray}
where 
\begin{equation}
{\tilde T}_{\mu \nu} \equiv -\frac{2}{\sqrt{-{\tilde g}}} \frac{\delta S_m}{\delta {\tilde g}^{\mu \nu}}
\end{equation}
is the energy-momentum tensor with respect to ${\tilde g}_{\mu \nu}$
and ${\tilde T} \equiv {\tilde g}^{\mu \nu}{\tilde T}_{\mu \nu}$.
Since all experiments are done with respect to ${\tilde g}_{\mu \nu}$, 
${\tilde T}_{\mu \nu}$ is the normal energy-momentum tensor we use
in the standard general relativity. 
For this reason, we call ${\tilde T}_{\mu \nu}$ physical energy-momentum tensor
\footnote{As it should be, it can be verified by explicit computation 
that the conservation law ${\tilde \nabla}_\mu {\tilde T}^\mu_{~\nu}=0$ is an
automatic consequence of
the combination of Eqs.~(\ref{eq-phi}) and (\ref{eq-g}).
}.
Since ${\tilde T}$ is independent of $\phi$,
the first EOM states that the effective potential for $\phi$ is 
\begin{equation}
V_{\rm eff}(\phi)=\frac{\mu^2}{2}\phi^2-\frac{1}{4}A^4(\phi) {\tilde T}, \label{effective-potential}
\end{equation}
for which we have $\Box \phi-V_{{\rm eff},\phi}=0$.

Now, let us choose the function $A^2(\phi)$ such that it is an 
even function and it monotonically decreases for $\phi >0$ and 
asymptotically approaches a constant value. 
One simple form that satisfies all these conditions is given by   
\begin{equation}
A^2(\phi)= 1-\varepsilon+\varepsilon e^{-\frac{\phi^2}{2M^2}},
\end{equation}
with $0 < \varepsilon < 1$.
Throughout this paper, we consider this form of $A(\phi)$ and 
assume that $\varepsilon ={\cal O}(1)$ 
and is not very close to $0$ nor $1$.
With this choice, the effective potential for $\phi$ in the presence
of non-relativistic matter, for which ${\tilde T}=-{\tilde \rho}$, becomes
\begin{equation}
V_{\rm eff}(\phi)=\frac{\mu^2}{2}\phi^2+
\frac{1}{4}{\left( 1-\varepsilon+\varepsilon e^{-\frac{\phi^2}{2M^2}}  \right)}^2 
{\tilde \rho}.
\end{equation}
The shape of $V_{\rm eff}$ is shown in the right panel of Fig.~\ref{fig:EP}.
When ${\tilde \rho}$ is uniform and the system is static, 
$\phi$ would take a constant value ${\bar \phi}$ which minimizes the effective potential.
Taylor-expanding $V_{\rm eff} (\phi)$ around $\phi=0$, we have
\begin{equation}
V_{\rm eff}(\phi)=\frac{1}{4} {\tilde \rho}+\frac{1}{2} 
\left( \mu^2-\frac{\varepsilon {\tilde \rho}}{2M^2} \right) \phi^2+{\cal O}(\phi^4). \label{Taylor-2nd}
\end{equation}
We find $\phi=0$ is stable for ${\tilde \rho} < \rho_{\rm PT} \equiv 2\mu^2 M^2/\varepsilon$,
but becomes unstable when ${\tilde \rho}$ exceeds $\rho_{\rm PT}$.
When ${\tilde \rho}>\rho_{\rm PT}$, ${\bar \phi}$ is given by
\begin{equation}
\frac{{\bar \phi}^2}{2M^2} =\ln f(\varepsilon,\rho_{\rm PT}/{\tilde \rho}),~~~~~
f(\varepsilon,\eta) \equiv \frac{2\varepsilon}{1-\varepsilon} 
{\left( \sqrt{1+\frac{4\varepsilon \eta}{{(1-\varepsilon)}^2}} -1
\right)}^{-1}. \label{phi-vev}
\end{equation}
We see that ${\bar \phi}$ depends only logarithmically on ${\tilde \rho}$.
Thus, unless ${\tilde \rho}$ takes extremely huge values, ${\bar \phi}$ 
is ${\cal O}(M)$.
To conclude, the scalar field undergoes spontaneous scalarization 
when ${\tilde \rho} > \rho_{\rm PT}$ is realized.

Let us next consider how the gravity behaves in the symmetric phase where 
${\bar \phi}=0$ is satisfied.
In the symmetric phase, the interactions between $\phi$ and
the other matter fields are, in the leading order, written as $\sim \phi^2 T_{\mu \nu}/M^2$.
As we will see later, for observationally interesting cases,
$M$ is typically much larger than ${\rm TeV}$ scale, {\it i.e.},
far beyond the energy scale accessible by any terrestrial experiments.
In this sense, $\phi$ completely decouples from the other matter fields
and behaves as a free massive scalar field.
Since $A^2 ({\bar \phi})=1$, assuming there is no excitation of $\phi$ field,
Eqs.~(\ref{eq-g}) reduce to the Einstein equations.
If $\phi$ is excited around $\phi=0$, excitation will be observed 
as dark component interacting only gravitationally with ordinary matter.
It is then natural to suppose that such excitation constitutes (a part of) dark matter.
More detailed analysis of this possibility including its production
mechanism will be discussed later.
Therefore, at low density region ${\tilde \rho} < \rho_{\rm PT}$,
GR is recovered.

\subsection{Gravity in spontaneous scalarization phase}
\label{subsec-gss}
Contrary to the symmetric phase, deviation from GR occurs
in the scalarization phase, which we will investigate in the following.
In the scalarization phase, 
due to a non-vanishing ${\bar \phi}$, matter fields interact with
$\phi$ with interaction strength proportional to ${\bar \phi}$.
This acts as a fifth force between matter fields.
Since $\phi$ is massive, the interaction range of the fifth force is limited
to $\sim 1/\mu$.
In addition to the emergence of the fifth force, 
field equations for gravity are also modified.
Assuming no excitation of $\phi$ field around ${\bar \phi}$,
Eqs.~(\ref{eq-g}), rewritten in terms of the Jordan-frame metric 
${\tilde g}_{\mu \nu}$,
become
\begin{equation}
{\tilde G}_{\mu \nu}+\Lambda_{\rm eff}{\tilde g}_{\mu \nu}=
8\pi G_{\rm eff} {\tilde T}_{\mu \nu}, \label{Jordan-metric-eq}
\end{equation}
where 
\begin{eqnarray}
&&\Lambda_{\rm eff}= 4\pi G_N \mu^2 {\bar \phi}^2 A^{-2}({\bar \phi})=
4\pi G_N \varepsilon \rho_{\rm PT} \ln f(\varepsilon,\rho_{\rm PT}/{\tilde \rho})
{\left( 1-\varepsilon+\frac{\varepsilon}{f(\varepsilon,\rho_{\rm PT}/{\tilde \rho})}
\right)}^{-1}, \label{SS-lambda}\\
&&G_{\rm eff} = A^2 ({\bar \phi}) G_N = 
\left( 1-\varepsilon+\frac{\varepsilon}{f(\varepsilon,\rho_{\rm PT}/{\tilde \rho})}
\right) G_N. \label{G-eff}
\end{eqnarray}
We find that ${\tilde g}_{\mu \nu}$ satisfies the Einstein equations
with the gravitational constant replaced by $G_{\rm eff}$ and
with the effective cosmological constant $\Lambda_{\rm eff}$.
Contrary to the case of the standard Higgs mechanism,
for which smaller cosmological constant is realized in the 
symmetry-breaking phase compared to that in the symmetric phase,
opposite phenomenon happens in the current model.
Namely, if there is no (or very tiny) cosmological constant
in the symmetric phase, then a positive vacuum energy of ${\cal O}(\rho_{\rm PT})$
emerges in the spontaneous scalarization phase.
Therefore, if the matter density is larger than $\rho_{\rm PT}$ 
but is still the same order of magnitude of $\rho_{\rm PT}$,
the effective cosmological constant will also play a non-negligible
role of gravitational physics. 
In the very high density region in which ${\tilde \rho} \gg \rho_{\rm PT}$,
we have
\begin{equation}
\Lambda_{\rm eff} \approx 4\pi G_N \frac{\varepsilon}{1-\varepsilon}
\rho_{\rm PT} \ln \left( (1-\varepsilon) \frac{\tilde \rho}{\rho_{\rm PT}} \right),~~~~~
G_{\rm eff} \approx (1-\varepsilon) G_N.
\end{equation}
We find that $\Lambda_{\rm eff}$ is enhanced only logarithmically from $\rho_{\rm PT}$.
Thus, in the very high density region, effect of the
effective cosmological constant is much smaller than the right
hand side of (\ref{Jordan-metric-eq}) and does not significantly
affect the dynamics.
The effective gravitational constant is reduced by $(1-\varepsilon)$.
Thus, gravity is weakened by this amount.

In the above argument, we have ignored the contribution of the scalar force
and focused only on the change in the pure gravity sector.
In order to evaluate the scalar force, let us consider a test point source
of its physical mass $M_S$ immersed in the static and uniform 
matter distribution in which spontaneous scalarization occurs. 
Presence of the point source distorts the scalar field from ${\bar \phi}$
by the amount $\delta \phi$ as well as the Einstein-frame metric from
$\eta_{\mu \nu}$ by the amount $h_{\mu \nu}$ \footnote{
For simplicity, we do not take into account the cosmological constant
term given by Eq.~(\ref{SS-lambda}) which exists in the background.
Inclusion of it is straightforward.
}.
We assume $M_S$ is so small that both $\delta \phi$ and $h_{\mu \nu}$
can be obtained by linear perturbation analysis.
Then, equation for $\delta \phi$ is obtained by linearizing Eq.~(\ref{eq-phi})
on the background $\phi={\bar \phi}$ given by Eq.~(\ref{phi-vev}).
On this background, we have
\begin{equation}
A^3 A_{,\phi} |_{\phi={\bar \phi}}
=-\frac{{\bar A}^2 \xi}{M},
\end{equation}
where ${\bar A} \equiv A({\bar \phi})$ and we have introduced
a dimensionless parameter $\xi$ defined by
\begin{equation}
\xi \equiv \frac{\varepsilon}{\sqrt{2}} 
\frac{\sqrt{\ln f(\varepsilon,\rho_{\rm PT}/{\tilde \rho})}}
{f(\varepsilon,\rho_{\rm PT}/{\tilde \rho})}. \label{def-alpha}
\end{equation}
Notice that in the deep scalarization phase for which 
${\tilde \rho} \gg \rho_{\rm PT}$,
this parameter is suppressed by a small factor $\rho_{\rm PT}/{\tilde \rho}$.
Since $\xi$ controls the coupling between the asymmetron
and matter fields,
the coupling is weak in the deep scalarization phase.

Using this quantity, the equation for $\delta \phi$ becomes
\begin{equation}
(\triangle -\mu^2) \delta \phi =-\frac{{\bar A}^2 \xi}{M}
{\tilde \rho_S},~~~~~~~~
{\tilde \rho_S}=\frac{M_S}{{\bar A}^3} \delta ({\vec x}).
\end{equation}
Solution of this equation is given by
\begin{equation}
\delta \phi (r)=\frac{\xi}{4\pi} \frac{M_S}{M} 
\frac{e^{-\mu r}}{{\bar A}r}.
\end{equation}
Metric perturbation $h_{\mu \nu}$ can be obtained in the standard manner.
Noting that the gravitational constant is ${\bar A}^2 G_N$ in the
scalarization phase and ${\bar A}r$ is the physical distance,
we have
\begin{equation}
h_{00}=2U,~~~~~h_{ij}=2U \delta_{ij},~~~~~~
U \equiv \frac{{\bar A}G_N M_S}{r},
\end{equation}
in the isotropic coordinates (or the PPN coordinates)
\footnote{If we are living in the scalarization phase,
we have to replace ${\bar A}^2 G_N$ by $G_N$ to satisfy the
local gravity experiments. See the last paragraph of the last 
section for relevant discussion.}.

The Jordan-frame metric with first order deviation from the background
is given by
\begin{equation}
{\tilde g}_{\mu \nu}=A^2(\phi)g_{\mu \nu}={\bar A}^2 \left( \eta_{\mu \nu}
+h_{\mu \nu}+{(\ln A^2)}_{,\phi} |_{\phi={\bar \phi}}~ \delta \phi 
\eta_{\mu \nu} \right).
\end{equation}
Since the constant overall factor ${\bar A}^2$ is irrelevant
to the following discussion, we will omit it.
Substituting the above results to ${\tilde g}_{\mu \nu}$, we find
\begin{eqnarray}
&&{\tilde g}_{00}=-1+2U+\frac{{\bar A}\xi^2 M_S}{2\pi M^2} \frac{e^{-\mu r}}{r}, \\
&&{\tilde g}_{ij}=\left( 1+2U - 
\frac{{\bar A}\xi^2 M_S}{2\pi M^2} \frac{e^{-\mu r}}{r} \right) \delta_{ij}.
\end{eqnarray}
We find that the scalar force described by the Yukawa potential
contributes to the metric perturbation in the Jordan-frame which
does not match the form predicted by the pure GR.
We can translate this contribution to the PPN parameter $\gamma$ 
(see, for instance, \cite{Will:2014kxa}).
This parameter is defined by 
${\tilde g}_{ij}=(1+2\gamma {\tilde U}) \delta_{ij}$ where
${\tilde U}$ is metric perturbation of the $00$ component, 
${\tilde g}_{00} =-1+2{\tilde U}$ (in GR, $\gamma =1$).
In the present case, $\gamma$ becomes
\begin{equation}
\gamma=1-\frac{2\lambda}{1+\lambda},~~~~~~~
\lambda \equiv \frac{\xi^2 e^{-\mu r}}{4\pi M^2 G_N}.
\end{equation}
Since $M$ appears in the denominator of $A_{,\phi}$
in Eq.~(\ref{eq-phi}),
naively one would expect that if $M$ is comparable or smaller than
the Planck scale $\sim G_N^{-1/2}$,
then the scalar force would become 
stronger than the gravitational force within the range $\sim \mu^{-1}$.
The above result shows that this naive expectation
is not correct since it is $\xi^2/(M^2 G_N)$ that determines
the magnitude of the deviation from GR.
As we mentioned earlier, $\xi$ becomes small in the 
deep scalarization phase and the system can become close to GR 
($|\gamma -1| \ll 1$) even when $M \lesssim G_N^{-1/2}$.

\subsection{Spontaneous scalarization only inside a compact object}
Having explained the basic picture of the spontaneous scalarization,
it is intriguing to analyze a situation where a dense object inside which
spontaneous scalarization occurs resides in vacuum.
To capture the essence of the phenomena, we make the following simplification 
that the object is static, uniform and spherically symmetric and
is made of non-relativistic matter and its size is much larger 
than the Schwarzschild radius so that metric in the Einstein-frame can be taken to
be the Minkowski one, but density is much larger than $\rho_{\rm PT}$.
These assumptions will be inappropriate in quantitative sense for dealing with
realistic astrophysical objects such as normal stars, white dwarfs, neutron stars and so on.
But we believe that the following result remains qualitatively correct.

\begin{figure}[tbp]
  \begin{center}
   \includegraphics[width=90mm]{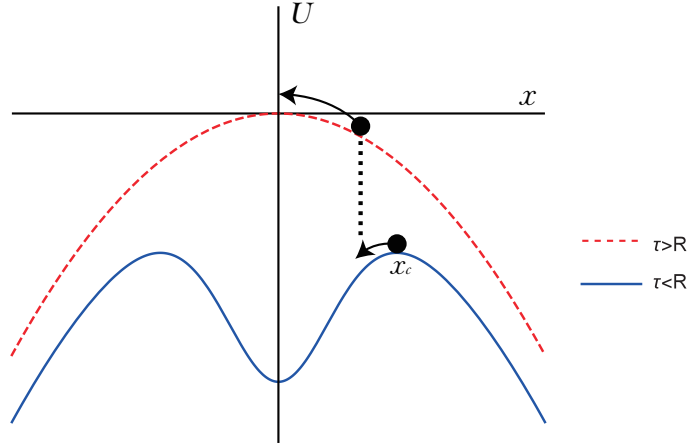}
  \end{center}
  \caption{Potential $U$ for $\tau<R$ and $\tau>R$. Initially,
  $x$ is at $x_c$ and asymptotically approaches zero for $\tau \to \infty$.}
  \label{fig:thin-shell}
\end{figure}

With the above assumptions, the equation for $\phi$ becomes
\begin{equation}
\frac{d^2 \phi}{dr^2}+\frac{2}{r} \frac{d\phi}{dr}-\frac{dV_{\rm eff}}{d\phi}=0.
\end{equation}
As is done in \cite{Khoury:2003rn}, 
let us perform the change of variables as
\begin{equation}
r \to \tau,~~~~~\phi \to x,~~~~~V_{\rm eff} \to -U.
\end{equation}
Then, the above equations becomes
\begin{equation}
\frac{d^2 x}{d\tau^2}+\frac{2}{\tau} \frac{dx}{d\tau}=-\frac{dU}{dx}, \label{eom-x}
\end{equation}
which represents motion of a point mass under the potential $U$ associated
with time dependent friction.
Denoting $R$ by the radius of the object,
$U$ changes its shape at $\tau=R$ as shown in Fig.~\ref{fig:thin-shell}.
What we want is a solution $x(\tau)$ with boundary condition,
\begin{equation}
x(0)=x_c,~~~{\dot x}(0)=0,~~~x(\tau \to \infty)=0.
\end{equation}
We follow \cite{Khoury:2003rn} to construct the approximate analytic solution for this kind of
problem.

When $R$ is large enough, $x$ stays near $x_c$ for a long time. 
This means that $x_c$ is very close to ${\bar \phi}$ at which $U'=0$.
At $\tau=R$, the friction had become negligible and the kinetic energy of $x$
is just enough to be compensated by the difference of potential energy
between $x=x(R)$ and $x=0$ so that $x$ asymptotically approaches zero.
Since the transition at $\tau=R$ happens near $x_c$, let us
replace $U(x)$ before $\tau=R$ by quadratic form around ${\bar \phi}$,
\begin{equation}
U(x) \approx -\frac{1}{2} m^2
{(x-{\bar \phi})}^2,
\end{equation}
where $m^2$ is the second derivative of $V_{\rm eff}$ evaluated
at ${\bar \phi}$,
\begin{equation}
m^2 \equiv V_{{\rm eff},\phi \phi}({\bar \phi})
=\frac{2 \mu^2}{\rho_{\rm PT}}
\left( 2\rho_{\rm PT}-(1-\varepsilon) 
\frac{\tilde \rho}{f(\varepsilon,\rho_{\rm PT}/{\tilde \rho})} \right)
\ln f(\varepsilon,\rho_{\rm PT}/{\tilde \rho}).
\end{equation}
Then, Eq.~(\ref{eom-x}) becomes
\begin{equation}
\frac{d^2 x}{d\tau^2}+\frac{2}{\tau} \frac{dx}{d\tau}=
m^2 (x-{\bar \phi}),
\end{equation}
whose solution with the initial condition $x(0)=x_c,~{\dot x}(0)=0$ is given by
\begin{equation}
x(\tau)={\bar \phi}+(x_c-{\bar \phi}) 
\frac{\sinh (m \tau)}{m \tau}.
\label{sol-x-inside}
\end{equation}
On the other hand, $x(\tau)$ for $\tau >R$ with boundary condition
$x \to 0$ for $\tau \to \infty$ is given by
\begin{equation}
x(\tau)=C \frac{e^{-\mu (\tau -R)}}{\tau}, \label{sol-x-outside}
\end{equation}
where $C$ is integration constant.
Requiring $x$ and ${\dot x}$ are continuous at $\tau=R$
determines $x_c$ and $C$ as
\begin{eqnarray}
&&x_c=\frac{-m (1+R\mu)+m \cosh (mR)+\mu \sinh (mR)}{m \cosh (mR)+\mu \sinh (mR)} {\bar \phi}, \\
&&C=\frac{m R \cosh (mR)-\sinh (mR)}{m \cosh (mR)+\mu \sinh (mR)} {\bar \phi}.
\end{eqnarray}
In the high density limit ${\tilde \rho} \gg \rho_{\rm PT}$,
$m^2$ becomes
\begin{equation}
m^2 = 2\mu^2 \ln \left( \frac{(1-\varepsilon) {\tilde \rho}}{\rho_{\rm PT}} \right),
\end{equation}
which is enhanced by log factor compared to $\mu^2$.
Then, neglecting $\mu$ terms in $x_c$ yields
\begin{equation}
x_c \approx \frac{-(1+R\mu)+\cosh (mR)}{\cosh (mR)} {\bar \phi}.
\end{equation}
Thus, if $R \gg m^{-1} (\gg \mu^{-1})$ is satisfied, then $\phi$
stays very close to ${\bar \phi}$ until the surface of the object
and then decays exponentially over length scale $\mu^{-1}$ outside
the object.
In other words, we can say that spontaneous scalarization 
occurs inside the object when the size of the object is much greater
than the Compton wavelength of $\phi$ in symmetric phase
(in addition to the trivial condition that density is greater
than $\rho_{\rm PT}$).

\subsection{Gravity outside the scalarized compact object}
Let us consider metric perturbation outside a compact object 
inside which spontaneous scalarization occurs.
As in the previous subsection, we assume that
the compact object is made of non-relativistic matter.
We assume that matter density is high enough so that spontaneous
scalarization occurs inside the object but not compact enough so 
that gravity is weak everywhere.
From Eqs.~(\ref{eq-g}), we see that this amounts to perform
perturbative expansion of the metric in the Einstein-frame 
around the Minkowski metric in terms of a dimensionless quantity
given by
(Schwarzshild radius)/(distance) \footnote{
One may wonder why we do not consider linear perturbation
in the Jordan-frame.
In order to see this is not feasible, let us express the field
equations (\ref{eq-g}) in terms of the Jordan-frame metric.
They are given by
\begin{equation}
{\tilde G}_{\mu \nu}+{\tilde g}_{\mu \nu} ({\tilde \nabla}^\alpha
\ln A {\tilde \nabla}_\alpha \ln A -2 {\tilde \nabla}^\alpha {\tilde \nabla}_\alpha \ln A)
+2 {\tilde \nabla}_\mu {\tilde \nabla}_\nu \ln A
=8\pi G_N \bigg[ -\left( \frac{1}{2} {\tilde \nabla}^\alpha \phi
{\tilde \nabla}_\alpha \phi +\frac{\mu^2}{2} \phi^2 \right) {\tilde g}_{\mu \nu}
+{\tilde \nabla}_\mu \phi {\tilde \nabla}_\nu \phi
+A^2 {\tilde T}_{\mu \nu} \bigg]. \nonumber
\end{equation}
We find that terms containing $\ln A$ on the left hand side of the above equation
are not associated with $G_N$.
This makes sense since they come from $G_{\mu \nu}$ for the Einstein-frame.
It is now clear that Jordan-frame metric cannot be expanded in terms
of (Schwarzshild radius)/(distance). 
Indeed, since $A$ changes by ${\cal O}(\varepsilon)$ from inside
to outside of the compact star, a large variation of the Jordan-frame metric
(exceeding ${\cal O}(G_N)$) is induced near the surface of the compact object. 
}.
In this subsection, we consider only first order correction
and treat the linearized Einstein equations.

We decompose the metric in the Einstein-frame as
\begin{equation}
g_{\mu \nu}=\eta_{\mu \nu}+h_{\mu \nu},~~~~~|h_{\mu \nu}| \ll 1,
\end{equation}
where $h_{\mu \nu}$ is proportional to $G_N$.
As usual, we introduce ${\bar h}_{\mu \nu}$ by
${\bar h}_{\mu \nu}=h_{\mu \nu}-\frac{1}{2} \eta_{\mu \nu}h$,
and impose a gauge condition $\partial^\mu {\bar h}_{\mu \nu}=0$.
Then the linearized field equations become
\begin{equation}
\Box {\bar h}_{\mu \nu}=-16\pi G_N A^2(\phi) {\tilde T}_{\mu \nu}.
\end{equation}
For the non-relativistic matter, we have
\begin{equation}
{\tilde T}_{\mu \nu} \approx {\rm diag} ( A^2 {\tilde \rho},~0,~0,~0).
\end{equation}
Thus, only the $t-t$ component becomes non-trivial,
\begin{equation}
\triangle {\bar h}_{00}=-16 \pi G_N A^4 {\tilde \rho}.
\end{equation}
Then, the gravitational potential $\Phi$ defined by $\Phi =-\frac{1}{2}h_{00}$
becomes
\begin{equation}
\Phi ({\vec x})=-G_N \int d^3x'~\frac{A^4 {\tilde \rho}({\vec x'})}{|{\vec x}-{\vec x'}|}.
\end{equation}
In particular, when ${\vec x}$ is very far from the object,
this becomes
\begin{equation}
\Phi ({\vec x}) \approx -\frac{G_N}{r} \int d^3x'~A^4 {\tilde \rho}({\vec x'}).
\label{g-potential}
\end{equation}
Since symmetry is restored ($\phi=0$) outside the star,
Einstein-frame is equivalent to the Jordan-frame in such a region.
Thus, the physical gravitational potential ${\tilde \Phi}$ is also given by
Eq.~(\ref{g-potential}).
The {\it distance} $r$ approaches the physical distance when $r$ is much larger
than the size of the star.
Noting that $A^3 d^3x$ is the physical volume element, physical mass $M_S$
is given by
\begin{equation}
M_S=\int d^3x~A^3 {\tilde \rho}({\vec x}). \label{physical-mass}
\end{equation}
If the size of the star is much bigger than $\mu^{-1}$,
spontaneous scalarization occurs inside the star and $\phi$ takes
the uniform value ${\bar \phi}$ given by Eq.~(\ref{phi-vev}) everywhere
inside the star except for the thin shell region near the surface.
Thus, it is reasonable to approximate $A$ to be uniform inside the star
($A=A({\bar \phi}) =A_{\rm in}$) and to have a step-function like 
transition at the surface of the star and to become unity outside the star.
With this simplification, we have
\begin{equation}
\Phi ({\vec x}) \approx -\frac{G_NA_{\rm in}M_S}{r}.
\end{equation}
For ${\tilde \rho} \gg \rho_{\rm PT}$, we have
$A_{\rm in} \approx \sqrt{1-\varepsilon}$.
Therefore, from the observer outside the star, $M$ appears to be
decreased by $A_{\rm in}$, or equivalently, $G_N$ appears to be decreased by $A_{\rm in}$.

Taking a component parallel to ${\tilde u}^\mu$ of the conservation law 
${\tilde \nabla}_\mu {\tilde T}^\mu_{~\nu}=0$ for the non-relativistic
matter and for the metric
${\tilde g}_{\mu \nu}=A^2 \eta_{\mu \nu}$,
we have
\begin{equation}
\frac{\partial}{\partial t} ( {\tilde \rho} A^3)+\frac{\partial}{\partial x^i}
({\tilde \rho} A^3 v^i)=0,
\end{equation}
where $v^i$ is defined by ${\tilde u}^i =v^i/A$.
Thus, the mass $M_S$ defined by Eq.~(\ref{physical-mass}),
which is the sum of mass of each particle that constitutes
the star, is conserved unless no matter escapes/enters the star.
This implies that gravitational potential far from the star
changes by $A_{\rm in}$ after the star undergoes the spontaneous scalarization.
At first glance, it appears that this conclusion is inconsistent with
the Birkhoff's theorem.
In order to understand this in more detail, let us consider a
spherically symmetric star whose density is initially smaller than $\rho_{\rm PT}$.
Let us assume that, at some time for some reason such as reduction of
the radiation pressure due to depletion of fuel to produce thermal energy, 
the star starts to shrink and the density eventually
exceeds $\rho_{\rm PT}$ before the star settles down to a new stable configuration.
By the time when the star becomes static again, 
the spontaneous scalarization is realized inside the star.
This final state is already described in previous subsections.
Let us write the Einstein-frame metric describing the transition as
\begin{equation}
ds^2=-(1+2 \Phi (t,r)) dt^2+(1+2\Lambda (t,r)) dr^2+r^2 d\Omega,
\end{equation}
where both $\Phi$ and $\Lambda$ are treated as linear perturbations
just as the previous subsection.
The scalar field also respects the spherical symmetry and hence $\phi=\phi(t,r)$.
Outside the star, the $t-r$ component of the Einstein equations (\ref{eq-g})
become
\begin{equation}
{\dot \Lambda}=4\pi G_N r {\dot \phi} \phi'.
\end{equation}
By integrating this equation along time with fixed $r$,
we have
\begin{equation}
r \big[ \Lambda (t \to \infty,r)-\Lambda (t\to -\infty, r) \big]
=4\pi G_N r^2 \int_{-\infty}^\infty dt~ {\dot \phi} \phi'.
\end{equation}
From the argument of the previous subsection,
the left hand side of the above equation is equal to 
$(A_{\rm in}-1)G_N M_S$ when $r$ is much bigger than the radius of the star. 
Thus, we have
\begin{equation}
(A_{\rm in}-1)M_S= \int_{-\infty}^\infty dt~ S_r{\dot \phi} \phi',
\end{equation}
where $S_r \equiv 4\pi r^2$ is the surface area of the sphere of radius $r$.
This result shows that change of the gravitational potential
before and after the spontaneous scalarization is compensated
by the emission of the scalar wave whose flux is given by ${\dot \phi} \phi'$.
Neglecting the metric perturbation, equation of motion for $\phi$
is given by
\begin{equation}
-{\ddot \phi}+\phi''+\frac{2}{r} \phi'+V_{{\rm eff},\phi}=0.
\end{equation}
Before the star starts to shrink, since the density of the star is less than $\rho_{\rm PT}$,
$\phi=0$ everywhere.
After the star starts to shrink and when the density exceeds $\rho_{\rm PT}$,
$V_{{\rm eff},\phi}$ at $\phi=0$ becomes unstable inside the star and
this acts as a force to push $\phi$ into the stable point.
In this way, $\phi$ inside the star changes its value \footnote{
Since scalarization with positive ${\bar \phi}$ and negative 
one are equally allowed,
scalarization occurs randomly on distance over the correlation length. 
As a result, the compact star just after the spontaneous scalarization
may be mixture of positive and negative ${\bar \phi}$ and two regions
are separated by domain wall.
Though this may lead to interesting phenomena, 
process of spontaneous scalarization with this effect
being taken into account is complicated and we do not consider
it in this paper.
}.
This change also excites the change of $\phi$ outside the star and
it propagates as a wave which decays as $\sim 1/r$.
Contribution of the scalar wave to the Jordan-frame metric far from the star
is given by
\begin{equation}
{\tilde h}_{\mu \nu} \supset -\varepsilon \frac{\phi^2}{2M^2} \eta_{\mu \nu} 
\propto r^{-2}. 
\end{equation}
Thus, this contribution is more suppressed for large $r$ compared to
gravitational potential and gravitational wave both of which decay as $\sim 1/r$
although the latter is absent in the present case from the beginning 
due to the simplified assumption that the system is spherically symmetric.
For a distant observer, the dominant deviation from GR caused by
the spontaneous scalarization is the change of the gravitational constant.

\section{Asymmetron as dark matter}
Having introduced a new scalar field $\phi$ which interacts with
standard matter only gravitationally in the symmetric phase, 
it is natural to identify it with dark matter.
As we will show, the spontaneous scalarization also provides a 
natural mechanism of fixing the abundance of dark matter within
the framework of primordial inflation.
  
Let us consider the effective potential during inflation.
Making the phenomenological approximation that inflation is caused by the fluid
with its equation of state ${\tilde P}_{\rm inf}=-{\tilde \rho}_{\rm inf}$,
we have
\begin{equation}
V_{\rm eff}(\phi)=\frac{\mu^2}{2}\phi^2
+{\left( 1-\varepsilon+\varepsilon e^{-\frac{\phi^2}{2M^2}} \right)}^2 {\tilde \rho}_{\rm inf}.
\end{equation}
True effective potential differs from this potential by the amount
of slow-roll parameters multiplied to the second term, which is 
small enough for our present purpose and we ignore it.
Due to the contribution of the pressure, the coefficient of the second term
on the right hand side is enhanced by a factor four compared to the
case of the non-relativistic matter.
As a result, ${\bar \phi}$ when the spontaneous scalarization occurs is given by
\begin{equation}
\frac{{\bar \phi}^2}{2M^2} =\ln f(\varepsilon,\rho_{\rm PT}/(4{\tilde \rho}_{\rm inf})), \label{phi-vev-inf}
\end{equation}
From this, we find that spontaneous scalarization occurs for 
${\tilde \rho}_{\rm inf} > \rho_{\rm PT}/4$,
which we assume to be satisfied.
The critical density is not equal to $\rho_{\rm PT}$ because
$\rho_{\rm PT}$ is defined as the critical density for the case of 
the non-relativistic matter (see below Eq.~(\ref{Taylor-2nd})).

During inflation, the $\phi$ field is fixed to the value given by Eq.~(\ref{phi-vev-inf}).
On top of this, since $\phi$ is almost massless during inflation $V_{{\rm eff, \phi \phi}} \ll H_{\rm inf}^2$,
perturbations originating from the quantum fluctuations of $\phi$ are generated.
We will come back to this issue later.  
After inflation, the Universe is reheated and dominated by radiation.
When this happens, ${\tilde T}$ vanishes and the effective potential reduces to
the bare potential \footnote{\label{footnote-1}Strictly speaking, this is not 
correct since there exists non-relativistic baryon component even in radiation 
dominated era after the QCD phase transition which occurs around temperature 
$T_{\rm QCD} \approx 200~{\rm MeV}$.
The baryon density at this temperature is estimated as 
$\rho_b (T_{\rm QCD}) \approx 6 \times 10^{-12}~{\rm GeV}^4$
for $\Omega_b = 0.04,~g_{s*}=20,~T_{\rm QCD}=200~{\rm MeV}$.
If $\rho_{\rm PT}$ is smaller than $\rho_b (T_{\rm QCD})$,
which is the case for $\mu < 10^{-11}~{\rm eV}$ (see Fig.~\ref{fig-mu.eps})
when we require asymmetron to be dark matter,
the baryon forces the asymmetron to undergo the spontaneous
scalarization at the time of the QCD phase transition.
As a result, the result (\ref{Omega-phi}) cannot be applied straightforwardly
and we need to modify it in an appropriate way.
Since $\rho_b (T_{\rm QCD})$ is much smaller than the nuclear density
which is an interesting target for $\rho_{\rm PT}$,
we do not consider this case in this paper and set $\mu > 10^{-11}~{\rm eV}$.
}.
As the Universe expands, the Hubble parameter gradually decreases and
at some point becomes equal to $\mu$.
Before this time, $\phi$ keeps its initial value fixed during inflation.
After this time, $\phi$ oscillates around the origin like $\phi (t) \sim \sin (\mu t)/t$
and $\rho_\phi$ behaves as non-relativistic matter.
Thus $\rho_\phi$ decreases as $1/a^3$ in the Einstein-frame ($a$ is the scale
factor in the Einstein-frame).
Then, the energy density of $\phi$ at present time is given by
\begin{equation}
\rho_{\phi,0}=\frac{1}{{(1+z_{\rm eq})}^3}
\frac{a_{\rm osc}^3}{a_{\rm eq}^3} \rho_{\phi,{\rm osc}}=
\frac{1}{{(1+z_{\rm eq})}^3} \frac{a_{\rm osc}^3}{a_{\rm eq}^3}
\frac{\mu^2}{2} {\bar \phi}^2 . 
\end{equation}
where the subscript ${\rm osc}$ in any quantity means that it is evaluated
when $\phi$ starts oscillations, {\it i.e.}, $\mu=H_{\rm osc}$ and
$a_{\rm eq}=1/(1+z_{\rm eq})$ is the scale factor at the time of 
matter radiation equality.
We assume that there is no additional entropy production after inflation.
Therefore, the entropy density of radiation decays as $1/{\tilde a}^3=A^3 a^3$
(${\tilde a}$ is the scale factor in the Jordan-frame).
With this assumption, we have
\begin{equation}
\rho_{\phi,0}=\frac{1}{{(1+z_{\rm eq})}^3}
\frac{g_{*s,{\rm eq}}}{g_{*s,{\rm osc}}}
{\left( \frac{g_{*,{\rm osc}}}{g_{*,{\rm eq}}} \right)}^{3/4}
{\left( \frac{\rho_{\rm r,eq}}{{\tilde \rho}_{\rm r,osc}} \right)}^{3/4} \frac{1}{A_{\rm inf}^3}
\frac{\mu^2}{2} {\bar \phi}^2,
\end{equation}
where $g_{*s},~g_*$ represents the effective degrees of freedom entering
in the entropy density, energy density of radiation, respectively.
By the time of matter-radiation equality, amplitude of $\phi$ has decreased enough
so that there is little difference between the Einstein-frame and the Jordan-frame,
{\it i.e.~,}$A_{\rm eq} \approx 1$ to a very good approximation.
Now, by using the Friedmann equation in the Einstein-frame
\begin{equation}
\mu^2=H_{\rm osc}^2=\frac{8\pi G_N}{3} A_{\rm inf}^4 {\tilde \rho}_{\rm r,osc},
\end{equation}
to eliminate ${\tilde \rho}_{\rm r,osc}$ and Eq.~(\ref{phi-vev-inf}) to eliminate ${\bar \phi}$,
we end up with
\begin{equation}
\Omega_{\phi,0} = \frac{\rho_{\phi,0}}{\rho_{c,0}}=
\frac{g_{*s,{\rm eq}}}{g_{*s,{\rm osc}}}
{\left( \frac{g_{*,{\rm osc}}}{g_{*,{\rm eq}}} \right)}^{3/4}
\frac{\varepsilon \rho_{\rm PT}}{2\rho_{c,0}} 
\ln f \left( \varepsilon,\rho_{\rm PT}/(4{\tilde \rho}_{\rm inf}) \right)
\frac{1}{A_{\rm inf}^2} {\left( \frac{H_0}{\mu} \right)}^{3/2} \Omega_{r,0}^{3/4}. \label{Omega-phi}
\end{equation}
If this quantity is equal to the observed $\Omega_{m,0}$,
then $\phi$ constitutes the whole dark matter. 
This requirement yields a relation between $\mu$ and $\rho_{\rm PT}$,
which is shown as $\mu_{\rm DM}=\mu_{\rm DM}(\rho_{\rm PT})$ in Fig.~\ref{fig-mu.eps}.
In this figure, the parameters are fixed as $g_{*,{\rm eq}}=g_{*s,{\rm eq}}=100$,
and $\varepsilon=1/2$.

Since there is strong upper limit on the deviation from general relativity
by the solar system experiments as well as terrestrial ones,
we require that the spontaneous scalarization occurs at density larger than
the Earth density, {\it i.e.,} $\rho_{\rm PT} \gg \rho_{\rm Earth} \approx 5 \times 10^{-17}~{\rm GeV}^4$
\footnote{It is possible that $\rho_{\rm PT} < \rho_{\rm Earth}$ and we are living
in the spontaneous scalarization phase.
One possibility is that $\varepsilon$ is very tiny $\varepsilon \ll 1$.
Since large deviation from GR never happens in any situation for such a case,
we do not consider this possibility in this paper.
The second possibility is that $\rho_{\rm PT}$ is the order of the critical
density of the Universe.
In this case, asymmetron behaves not as dark matter but as dark energy.
We will briefly discuss this scenario in the last section.}.
Combining this with the footnote \ref{footnote-1},
our primary interest for $\rho_{\rm PT}$ is $\rho_{\rm PT} \gtrsim 10^{-11} {\rm GeV}^4$.
Then, from Fig.~\ref{fig-mu.eps}, we find that the corresponding restriction on $\mu$
is given by $\mu \gtrsim 10^{-11}~{\rm eV}$ which we regard as the possible minimum value
of our interest.

\subsection{Isocurvature constraint}
We saw in the previous subsection that spontaneous scalarization occurs during
the primordial inflation and this provides a mechanism of preparing non-zero value
of the asymmetron to realize its coherent oscillations which behave 
as non-relativistic matter interacting only gravitationally with other matter fields.
There is indeed a parameter range of $\mu$ and $M$ where the energy density of
asymmetron is equal to that of dark matter.
But before asymmetron can be considered as a candidate of dark matter, 
it must satisfy other observational constraints.
There are two non-trivial observational constraints, which we will consider below.

The first constraint is the non-detection of the dark matter isocurvature perturbation.
Since the $\phi$ field is almost massless during inflation, 
this field acquires almost scale-invariant classical fluctuations during inflation 
when each wavelength mode crosses the Hubble horizon.
In addition to this, the standard adiabatic perturbations are also generated 
from classical fluctuations of either inflaton or other light fields,
which are equally shared by all the existing particle species such as photons,
baryons and dark matter.
On top of this, dark matter has its own fluctuations coming from the fluctuations
of the $\phi$ field itself and these fluctuations act as isocurvature perturbations
having no correlation with the adiabatic ones.
Since there is a strong upper limit on the amplitude of the isocurvature perturbations
imposed by CMB observations,
this limit can be converted to the constraint on the domain of the 
$(\mu,~\rho_{\rm PT})$ plane.
To see this in more quantitative manner,
let us first introduce the dark matter isocurvature perturbation ${\cal S}_{\rm DM}$ by \cite{Hinshaw:2012aka}
\begin{equation}
{\cal S}_{\rm DM}=\frac{\delta \rho_{\rm DM}}{\rho_{\rm DM}}-\frac{3}{4} \frac{\delta \rho_\gamma}{\rho_\gamma},
\end{equation}
where $\delta \rho_\gamma$ is the density perturbation of photons.
This quantity is conserved as long as the scale considered is super-Hubble scale. 
In the present case, $\rho_{\rm DM}=\rho_\phi$.

As is done in \cite{Sasaki:2006kq} (but for a different model),
we adopt the approximation of the sudden transition that $\phi$ is completely frozen before
$H=\mu$ (hence $\rho_\phi$ is constant in time) and starts to oscillate 
exactly when $H=\mu$ and 
behaves as non-relativistic matter $\rho_\phi \propto a^{-3}$ \cite{Turner:1983he}.
Then, the hypersurface on which $\phi$ starts to oscillate coincides 
with the one with constant total energy density.
Since ${\cal S}_{\rm DM}$ is independent of the choice of time slicing, 
we can compute $\delta_r$ and $\delta_\phi$ in any time slicing and 
we take the $H=m$ hypersurface for this purpose.
On this hypersurface, we have
\begin{equation}
\rho_r ({\vec x})+\rho_\phi ({\vec x})=\rho_{\rm tot}=\frac{3\mu^2}{8\pi G_N}.
\end{equation}
Decomposing this relation into the background part and perturbation part and
extracting the perturbation part, we have
\begin{equation}
\delta_r ({\vec x})=-\frac{\Omega_\phi}{1-\Omega_\phi} \delta_\phi ({\vec x}),
\end{equation}
where $\Omega_\phi = \rho_\phi/\rho_{\rm tot}$, evaluated at time when $H=\mu$.
Plugging this relation into the definition of ${\cal S}_{\rm DM}$, we have
\begin{equation}
{\cal S}_{\rm DM} ({\vec x})=\left( 1+\frac{3}{4} \frac{\Omega_\phi}{1-\Omega_\phi} \right)
\delta_\phi ({\vec x}) \approx \delta _\phi ({\vec x}),
\end{equation}
where we have used $\Omega_\phi \ll 1$ since the time when
asymmetron starts to oscillate for the range of $\mu$ of our interest
is much earlier than the time of matter-radiation equality.
Since $\rho_\phi =\mu^2 \phi^2/2$ in the radiation dominated era, 
we finally have
\begin{equation}
{\cal S}_{\rm DM} ({\vec x})=\frac{2 \delta \phi}{\bar \phi}.
\end{equation}
Here, $\delta \phi$ is the perturbation quantum mechanically generated during inflation.
This is uncorrelated with the (adiabatic) curvature perturbation which is sourced
by other fields.
The corresponding power spectrum of ${\cal S}_{\rm DM}$ is given by
\begin{equation}
{\cal P}_{\rm CDM}=\frac{4}{{\bar \phi}^2} 
{\left( \frac{H_{\rm inf}}{2\pi} \right)}^2
=\frac{8G_N \mu^2}{3\pi} A_{\rm inf}^4 
\frac{{\tilde \rho}_{\rm inf}}{\varepsilon \rho_{\rm PT}}
\frac{1}{\ln f(\varepsilon,\rho_{\rm PT}/(4{\tilde \rho}_{\rm inf}))}, \label{P-iso}
\end{equation}
where the modified Friedmann equation 
\begin{equation}
3H_{\rm inf}^2=8\pi G_N A^4 ({\bar \phi}) \rho_{\rm inf},
\end{equation}
is used to obtain the final expression.
The upper limit on the uncorrelated dark matter isocurvature perturbation
by WMAP 9yr is given by \cite{Hinshaw:2012aka}
\begin{equation}
\frac{{\cal P}_{\rm CDM}}{{\cal P}_{\cal R}} < \frac{\alpha}{1-\alpha},~~~~~\alpha < 0.047~~
(95 \%~CL),
\end{equation}
where ${\cal P}_{\cal R}$ is the power spectrum of the adiabatic perturbation.
For fixed $\rho_{\rm PT}$ and ${\tilde \rho}_{\rm inf}$,
this bound can be converted into the upper bound on $\mu$ which is shown
as a line of $\mu=\mu_{\rm iso}$ in Fig.~\ref{fig-mu.eps}.
We find that a line $\mu=\mu_{\rm DM}$ lies above $\mu=\mu_{\rm iso}$ for the case of
high energy inflation scale such as ${\tilde \rho}_{\rm inf}={(10^{15}~{\rm GeV})}^4$.
Thus, the scenario of asymmetron being 
responsible for the total content of dark matter
is incompatible 
with those inflation models where the energy scale is 
as high as $\sim 10^{15}~{\rm GeV}$.
The line $\mu=\mu_{\rm DM}$ comes below the isocurvature constraint line 
if the inflation energy scale is lowered, which can be understood from the expression of 
${\cal P}_{\rm DM}$ given by Eq.~(\ref{P-iso}).
The equation shows that ${\cal P}_{\rm DM}$ is basically proportional to ${\tilde \rho}_{\rm inf}$
(the denominator depends only logarithmically on ${\tilde \rho}_{\rm inf}$).
Indeed, if ${\rho}_{\rm inf}$ is as low as $10^{13}~{\rm GeV}$, 
the isocurvature constraint is evaded for all the range of $\rho_{\rm PT}$ we are interested in
({\it i.e.~} nuclear energy density).

\subsection{Constraint from the fifth force experiments}
As is already mentioned, we are interested in the case where $\rho_{\rm PT}$ is between the
stellar density and the nuclear density realized in the neutron stars
so that spontaneous scalarization takes place in compact astrophysical objects
and ${\cal O}(\varepsilon )$ deviation from general relativity occurs only in such regions.
In the asymmetron model, the solar system is in the symmetric phase (${\bar \phi}=0$).
As we saw in the previous section, GR is exactly recovered in this phase
and this model passes the solar system and terrestrial experiments that 
have placed very strong limit on deviation from GR.
However, this conclusion must be reconsidered more carefully if we take the scenario
of the asymmetron being dark matter.
In this case, $\phi$ is coherently oscillating in time with angular frequency $\mu$,
which describes the cold dark matter.
The value of $\phi$ averaged over time longer than the oscillation period is zero,
but the value at each time is different from zero.
Therefore, the assumption of no excitation of $\phi$ in the symmetric phase is violated
if we require $\phi$ to be dark matter.

\begin{figure}[tbp]
  \begin{center}
  \subfigure[${\tilde \rho}_{\rm inf}^{1/4}=10^{13}~{\rm GeV}$]{
   \includegraphics[width=90mm]{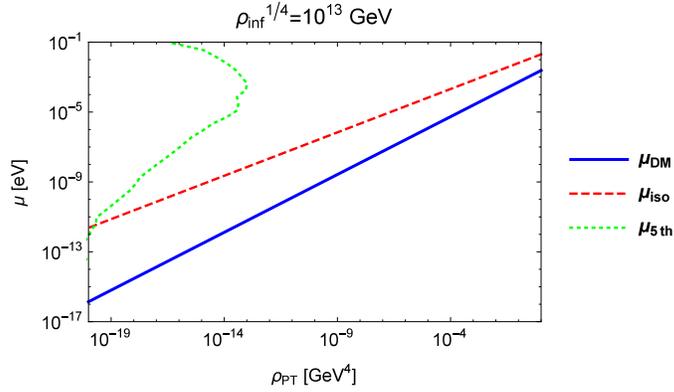}}
  \subfigure[${\tilde \rho}_{\rm inf}^{1/4}=10^{15}~{\rm GeV}$]{
   \includegraphics[width=90mm]{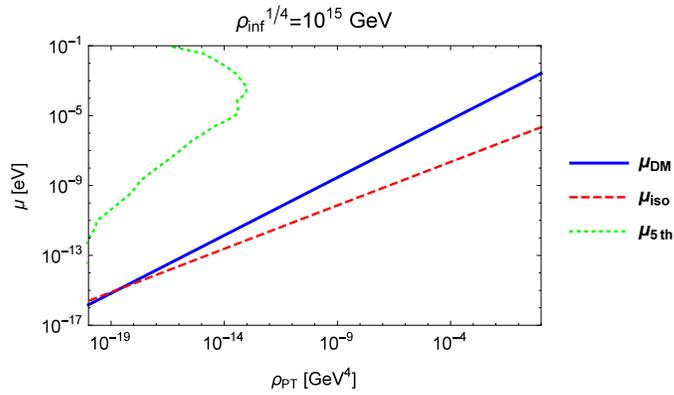}}  
   \caption{Three curves in each panel are $\mu_{\rm DM}$, $\mu_{\rm iso}$
   and $\mu_{\rm 5th}$. Right to each curve of $\mu_{\rm iso}$ and $\mu_{\rm 5th}$ 
   is the allowed region satisfying the observational constraints.}
   \label{fig-mu.eps}
  \end{center}
\end{figure}

When $\phi$ is oscillating in the symmetric phase, 
the interaction between matter fields and $\phi$ field given by
$A^3 A_\phi {\tilde T}$ (see Eq.~(\ref{eq-phi})) also oscillates
as $- \varepsilon a \sin (\mu t)/M^2$ ($A \approx 1$ is assumed),
where $a$ is the amplitude of the oscillations of $\phi$ 
and is given by $a^2 = 2 \langle \phi^2 \rangle$ ($\langle \cdots \rangle$ 
represents the time average over oscillation period).
The amplitude $a$ is determined from the requirement that $\rho_\phi = \mu^2 \langle \phi^2 \rangle$
coincides with the dark matter density. This condition yields
\begin{equation}
a^2 = \frac{2 \rho_{\rm DM}}{\mu^2}. \label{eq-a2}
\end{equation}
The effect of oscillating $\phi$ can be observed as the periodically 
time varying gravitational force (period is $\pi / \mu$) acting on two massive bodies 
with maximum given by $\sim \varepsilon a/M^2$ on top of the standard gravitational force.
In order to see this, let us determine the gravitational potential in the Jordan-frame.
To simplify the analysis, based on the fact that the time scale of
interest ({\it e.~g.}, time duration of experiments) is much larger than the oscillation
period $2\pi /\mu$;
\begin{equation}
\frac{2\pi}{\mu} \approx 4\times 10^{-4}~s~
{\left( \frac{\mu}{10^{-11}~{\rm eV}} \right)}^{-1},
\end{equation}
for a range of interest $\mu \gtrsim 10^{-11}~{\rm eV}$,
we make the approximation that only the averaged value of $\phi$ enters the 
measurable gravitational potential and asymmetron both of which
are simultaneously generated by the source object such as the Earth.
Denoting $\delta \phi$ by the small deviation from the background $\phi$ caused by
the presence of the point mass with its mass $M_s$ (${\tilde \rho}=M_s \delta ({\vec x})$),
equation for $\delta \phi$ is obtained by linearizing Eq.~(\ref{eq-phi});
\begin{equation}
\triangle \delta \phi-\mu^2 \delta \phi+\frac{\varepsilon \sqrt{\langle \phi^2 \rangle}}{2M^2} {\tilde \rho}=0.
\end{equation}
Solution of this equation is given by
\begin{equation}
\delta \phi (r)=\frac{\varepsilon \sqrt{\langle \phi^2 \rangle}}{8\pi M^2} \frac{M_s}{r}e^{-\mu r}.
\end{equation}
As a result, time-time component of the Jordan-frame metric is given by
\begin{equation}
{\tilde g}_{00}=-1+\frac{2G_NM_s}{r} F(r),~~~~~F(r) \equiv 1+\frac{\varepsilon^2 \langle \phi^2 \rangle}{16\pi M^4 G_N} 
e^{-\mu r}. \label{eq-g00}
\end{equation}
The function $F(r)$ represents the modification from the standard gravitational potential\footnote{
In addition to the modification by $F(r)$, change of the Newton's constant caused by
the change of $A$ ($\langle A^2 \rangle \neq 1$) also modifies the gravitational.
However, this correction is negligibly small and we do not take this effect into account in Eq.~(\ref{eq-g00})}.
Eliminating $\langle \phi^2 \rangle$ by Eq.~(\ref{eq-a2}), we have
\begin{equation}
F(r)=1+\frac{\mu^2}{4\pi G_N \varepsilon} \frac{\rho_{\rm DM}}{\rho_{\rm PT}^2}
e^{-\mu r}.
\end{equation}
As expected, the $\phi$ field contributes to the force between two
bodies as Yukawa type force.
Various experiments have been performed to test the inverse square law of
gravity. One of the typical modifications of the inverse square law
which is actively tested by experiments is exactly the form given by $F(r)$.
In \cite{Adelberger:2003zx}, deviation from the inverse square law of
a form $F(r)=1+\alpha e^{-r/\lambda}$ is assumed and upper limit on $\alpha$
is summarized for a wide range of $\lambda$ from $10^{-9}~{\rm m}$
to $10^{15}~{\rm m}$.
We converted this constraint in $\lambda - \alpha$ plane to the
constraint in $\rho_{\rm PT} - \mu$ plane.
The result is shown as a green(dotted) curve in Fig.~\ref{fig-mu.eps}.
The region satisfying the fifth force experiments is right to the 
green(dotted) curve.
We find that the constraint from the fifth force experiment is much
weaker than the isocurvature one and is always satisfied for any 
interesting range of $\rho_{\rm PT}$.

\section{Discussion and conclusion}
We have proposed the asymmetron model, a class of scalar-tensor
theories, in which the significant deviation from GR occurs only
in high matter density region.
This is an extended version of the Damour-Esposito model proposed
in \cite{Damour:1993hw} by adding mass term and allowing the 
energy scale appearing in the conformal factor to differ
from the Planck scale.
We have shown that the asymmetron model can be consistently
embedded in the cosmological framework.
In particular, spontaneous scalarization caused by inflaton in
a dynamical way provides
the initial condition for the subsequent coherent oscillations of the asymmetron.
The damped oscillation has nice properties in that it not only makes the 
asymmetron behave as cold dark matter but also makes GR a cosmological attractor.
Oscillating asymmetron yields periodically varying fifth force but its magnitude
is far below the current experimental sensitivities and the model we studied
in this paper is in practice indistinguishable from
GR in the present Universe except inside
dense compact objects and easily passes the solar-system and terrestrial experiments.
There is a range of parameter space where the asymmetron can 
saturate the whole dark matter component
and at the same time significant deviation from GR
in the present Universe occurs only inside the dense compact objects such as neutron stars.

In the spontaneous scalarization phase, the gravitational constant becomes
smaller than that in the symmetric phase, namely
the value determined in laboratories.
Thus, the gravity is weakened only inside dense compact objects,
which is a dominant modification from GR.
The scalar force also operates among matter with strength 
given by $\sim \xi^2/(M^2 G_N)$ (see Sec.~\ref{subsec-gss}) 
compared with gravitational force.
In the deep scalarization phase where $\xi \ll 1$, 
the scalar force can become tiny even when $M$ is smaller
than the Planck mass $G_N^{-1/2}$.
Furthermore, the interaction range is limited by $\sim \mu^{-1}$
and for the range of our interest this scale is rather short.
As a result, weakening of gravity is the dominant feature representing
deviation from GR when the density of compact object is much bigger than
the critical density above which spontaneous scalarization occurs.
This suggests that the size of the compact star in the asymmetron model becomes 
larger than that in GR.
Since the Chandrasekhar mass is proportional to $G_N^{-3/2}$,
we expect that the Chandrasekhar mass in our model should be larger
than that in GR for compact stars undergoing spontaneous scalarization.

There are many issues that we did not consider in this paper and deserve further investigations.
In this paper, we mainly focused on the mechanism of the spontaneous scalarization
in the asymmetron model, its basic properties and embedding it in the cosmological framework.
Obviously, the next thing to do is to investigate how to test this model in astrophysics,
in particular in connection to gravitational wave observations. 
In this context, it is first of all interesting to clarify how the stellar structure 
(such as mass-radius relation and the Chandrasekhar mass) 
in the asymmetron model is modified from GR.
Gravitational waves from compact binaries are the main target of the laser interferometers.
Dynamics of binaries, waveform of the gravitational waves and detectability by
using the interferometers for the original DEF model have been studied \cite{Barausse:2012da,Doneva:2013qva,Shibata:2013pra,Palenzuela:2013hsa,Sampson:2014qqa,Sotani:2014tua,Pani:2014jra,Taniguchi:2014fqa,Silva:2014ora,Ponce:2014hha}.
Performing the similar analysis for the asymmetron model will help to elucidate 
what observation is the best probe for exploring the asymmetron model.

Another intriguing thought is the possibility of asymmetron being responsible
for dark energy.
In this paper, we have focused on the case where $\rho_{\rm PT} \gg \rho_{\rm Earth}$
and the spontaneous scalarization occurs only in extremely high density region.
On the other hand, if $\rho_{\rm PT}$ is the order of the current critical density of
the Universe, we expect that the scalarization persists
until present epoch and the mass term $\frac{1}{2}\mu^2 \phi^2$ approximately
plays the role of the cosmological constant.
Taking $M$ to be Planck mass, this is achieved if $\mu$ is chosen to
be around the Hubble constant $H_0$.
This means that locally, such as in the solar system, asymmetron
mediates a long range force in addition to the gravitational force.
However, since the solar system is in the deep scalarization phase,
the scalar force is suppressed by the factor $\xi$ given by Eq.~(\ref{def-alpha}).
Indeed, if we take $\rho_{\rm PT}$ to be the present critical
density of the Universe and ${\tilde \rho}$ the density of
solar wind (we assume one proton per cubic centimeter) and
$\varepsilon = 1/2$, we have $\xi \approx 5 \times 10^{-5}$
and $1-\gamma \approx 2 \times 10^{-10}$.
This value is much below the current constraint $|\gamma-1| \lesssim 10^{-5}$ 
obtained from the time delay measurement \cite{Will:2014kxa}.
Thus, the asymmetron as dark energy can safely pass the
solar system constraints.
The most characteristic feature would be time-dependence of the
gravitational constant (see Eq.~(\ref{G-eff})) through the time-dependence
of the matter density ${\tilde \rho}$ due to the cosmic expansion.
On cosmological scales, the gravitational constant gradually 
increases as the Universe expands and it is interesting to
investigate how the large scale structure is affected by such
a time varying gravitational constant.

\section*{Acknowledgments}
We would like to thank Toshikazu Shigeyama for useful comments.
P.C. wishes to acknowledge the hospitality during his stay at RESCEU
as a visiting professor.
This work was supported by JSPS
Grant-in-Aid for Young Scientists (B) No.15K17632
(T.S.), MEXT Grant-in-Aid for Scientific Research on
Innovative Areas New Developments in Astrophysics
Through Multi-Messenger Observations of Gravitational
Wave Sources No.15H00777 (T.S.), JSPS KAKENHI No. 15H05888 (T.S. and J.Y.)
and 15H02082(JY).

\bibliographystyle{unsrt}
\bibliography{draft}

\end{document}